\documentclass[10pt,preprint2]{emulateapj}
\usepackage{amsmath}
\usepackage{float}

\slugcomment{Received 2015 April 2015 - Accepted 2015 July 24}

\shorttitle{Environment of the massive starburst \textit{HFLS3}}
\shortauthors{Laporte et al.}

\begin{document}


\title{Environment of the submillimeter-bright massive starburst HFLS3 at $z$$\sim$6.34}


\author{N. Laporte$^{1,2,3}$, I. P\'erez-Fournon$^{1,2}$,  J. A. Calanog$^{4}$, A. Cooray$^{4,5}$, J.L. Wardlow$^6$, 
J. Bock$^{7,8}$, C. Bridge$^5$, D. Burgarella$^9$, R. S. Bussmann$^{10}$, A. Cabrera-Lavers$^{1,2,21}$, C.M. Casey$^4$, D. L. Clements$^{11}$, A. Conley$^{12}$, H. Dannerbauer$^{25}$, D. Farrah$^{13}$, H. Fu$^{14}$, R. Gavazzi$^{15}$, E. A. Gonz\'alez-Solares$^{22}$, R. J. Ivison$^{16,17}$, B. Lo Faro$^{9}$, B. Ma$^{4}$, G. Magdis$^{18}$, R. Marques-Chaves$^{1,2}$, P. Mart\'inez-Navajas$^{1,2}$, S. J. Oliver$^{19}$, W. A. Osage$^{4}$, D. Riechers$^{10}$, D. Rigopoulou$^{18,24}$, D. Scott$^{20}$, A. Streblyanska$^{1,2}$, J.D. Vieira$^{5,23}$}



\altaffiltext{1}{Instituto de Astrofisica de Canarias (IAC), E-38200 La Laguna, Tenerife, Spain.}
\altaffiltext{2}{Departamento de Astrofisica, Universidad de La Laguna (ULL), E-38205 La Laguna, Tenerife, Spain}
\altaffiltext{3}{Instituto de Astrof\'{\i}sica, Facultad de F\'{i}sica, Pontificia Universidad Cat\'{o}lica de Chile, 306, Santiago 22, Chile}
\altaffiltext{4}{Dept. of Physics \& Astronomy, University of California, Irvine, CA 92697}
\altaffiltext{5}{California Institute of Technology, 1200 E. California Blvd., Pasadena, CA 91125, USA}
\altaffiltext{6}{Dark Cosmology Centre, Niels Bohr Institute, University of Copenhagen, Juliane Maries Vej 30, 2100 Copenhagen, Denmark}
\altaffiltext{7}{California Institute of Technology, 1200 E. California Blvd., Pasadena, CA 91125, USA}
\altaffiltext{8}{Jet Propulsion Laboratory, 4800 Oak Grove Drive, Pasadena, CA, 91109, USA}
\altaffiltext{9}{Laboratoire d'Astrophysique de Marseille, Aix-Marseille University, CNRS, 13013 Marseille, France}
\altaffiltext{10}{Department of Astronomy, Space Science Building, Cornell University, Ithaca, NY, 14853-6801, USA}
\altaffiltext{11}{Astrophysics Group, Imperial College London, Blackett Laboratory, Prince Consort Road, London SW7 2AZ, UK}
\altaffiltext{12}{Dept. of Astrophysical and Planetary Sciences, CASA 389-UCB, University of Colorado, Boulder, CO 80309, USA}
\altaffiltext{13}{Departement of Physics, Virginia Tech., Blacksburg, VA 24061, USA}
\altaffiltext{14}{Departement of Physics \& Astronomy, University of Iowa, Iowa City, IA 52242, USA}
\altaffiltext{15}{Institut d'Astrophysique de Paris, UMR 7095, CNRS, UPMC, Univ. Paris 06, 98bis Blvd. Arago, F-75014 Paris, France}
\altaffiltext{16}{European Southern Observatory, Karl Schwarzschild Strasse 2, D-85748 Garching, Germany}
\altaffiltext{17}{Institute for Astronomy, University of Edinburgh, Royal Observatory, Blackford Hill, Edinburgh EH9 3HJ, UK}
\altaffiltext{18}{Department of Astrophysics, Denys Wilkinson Building, University of Oxford, Keble Road, Oxford OX1 3RH, UK}
\altaffiltext{19}{Astronomy Centre, Dept. of Physics \& Astronomy, University of Sussex, Brighton BN1 9QH, UK}
\altaffiltext{20}{Department of Physics \& Astronomy, University of British Columbia, 6224 Agricultural Road, Vancouver, BC V6T 1Z1, Canada}
\altaffiltext{21}{GTC Project, E-38205 La Laguna, Tenerife, Spain }
\altaffiltext{22}{Institute of Astronomy, Cambridge University, Madingley Road, Cambridge CB3 OHA, UK }
\altaffiltext{23}{Department of Astronomy and Department of Physics, University of Illinois, 1002 West Green Street, Urbana, IL 61801, USA}
\altaffiltext{24}{RAL Space, Science, and Technology Facilities Council, Rutherford Appleton Laboratory, Didcot OX11 0QX, UK}
\altaffiltext{25}{Institut fur Astronomie, Universitat   Wien,Turkenschanzstrasse 17, A-1160 Wien, Austria}


\begin{abstract}
We describe the search for Lyman-break galaxies (LBGs) near the sub-millimeter bright starburst galaxy 
\textit{HFLS3} at $z$$=$6.34 and a study on the environment of this massive galaxy during the end of reionization.
We performed two independent selections of LBGs on images obtained with the \textit{Gran Telescopio Canarias} (GTC) and the \textit{Hubble Space Telescope} (HST)
by  combining non-detections in bands blueward of the Lyman-break and color selection. 
A total of 10 objects fulfilling the LBG selection criteria at $z$$>$5.5 were selected over the 4.54  and 55.5 arcmin$^2$ covered by our HST and GTC images, respectively. 
The photometric redshift, UV luminosity, and the star-formation rate of these sources were estimated with models of their spectral energy distribution.
These $z$$\sim$6 candidates have physical properties and number densities in agreement with previous results. 
The UV luminosity function at $z$$\sim$6  and a Voronoi tessellation analysis of this field shows no strong evidence for an overdensity of relatively bright objects (m$_{F105W}$$<$25.9) associated with \textit{HFLS3}. However, the over-density parameter deduced from this field and the surface density of objects can not excluded definitively the LBG over-density hypothesis. Moreover we identified three faint objects at less than three arcseconds from \textit{HFLS3}
with color consistent with those expected for $z$$\sim$6 galaxies. Deeper data are needed to confirm their redshifts and to study their association with
\textit{HFLS3} and the galaxy merger that may be responsible for the massive starburst.
\end{abstract}





\section*{Introduction}
One of the most important questions of modern astronomy is undoubtedly the formation and evolution of the first luminous objects in our Universe. During the last decade, the arrival of new facilities with capabilities to push even further the boundaries of our Universe, such as for example \textit{ Herschel Space Observatory} \citep{HSO}, the HST/WFC3 \citep{WFC3}, and VLT/HAWK-I \citep{HAWKI}, contributed to considerable advances in our understanding of the first billion years.  
The number of $z>$6 galaxies presently confirmed by spectroscopy has strongly increased over the last few years (e.g.  \citealt{Schenker12}, \citealt{Vanzella11} ) allowing to better constrain their physical properties in terms of the star-formation rate (SFR), stellar mass, and reddening (e.g. \citealt{deBarros12}). The evolution of such galaxies is now relatively well-constrained and supported by spectroscopic observations out to $z$$\sim$6 (e.g. \citealt{LeFevre14}), and photometric studies are now slowly starting to give robust constraints of galaxies out to $z\sim$10 (e.g. \citealt{Bouwens14}).

The detection of large numbers of dusty, massive starburst galaxies at $z\sim$2 was a surprise when they were first identified in 1997 (\citealt{Smail97}; \citealt{Hughes98}; for a review see \citealt{Casey14}). Theoretically the existence of dusty, massive starbursts at such early epochs is difficult to explain and thorough observational constraints on their properties provide a stringent test of the galaxy formation models (e.g. \citealt{Baugh05}). The \textit{HerMES} survey \citep{Hermes} has identified a population of dusty starburst galaxies at $z \ge$ 4 \citep{Dowell14} that are not predicted by today's galaxy formation paradigms (e.g. \citealt{Hayward12}). The efficient identification and detailed study of these rare $z\geq$ 4 star-bursting galaxies is important for understanding their progenitor populations, drivers of cold gas accretion \citep{Carilli10}, and their descendants. It has been shown by previous studies that several massive starburst at high-$z$ could be proto-cluster members, but none of them has been identified at $z\ge$6 (\citealt{Daddi09}, \citealt{Capak11}).

Recently, \citet{Riechers13} have discovered an extreme sub-millimeter starburst at $z=$6.34, called \textit{HFLS3} hereafter, over 21 deg$^2$ of the \textit{Herschel}/SPIRE \citep{SPIRE} data from the HerMES survey. This object is lensed by a two-component galaxy system at $z\sim$2.1 involving a magnification of 2.2 $\pm$0.3 \citep{Cooray14}. Its physical properties based on the rest-frame UV emission results in a SFR  of 1320 M$_{\odot}$yr$^{-1}$ and dust and stellar masses of 3 $\times 10^{8}$ M$_{\odot}$ and 5 $\times 10^{10}$ M$_{\odot}$, respectively, making this source one of the most massive starburst presently known during the epoch of reionization (EoR).

A recent paper aiming to detect sub-millimeter emission in the vicinity of \textit{HFLS3} with SCUBA-2 data \citep{Robson14} found no evidence in favor of an over-density of dusty galaxies associated with \textit{HFLS3}. In this paper, we present results of an optical and near-IR analysis of the environment of \textit{HFLS3} aiming to recover Lyman-break galaxies at $z$$\sim$6. We combine two datasets: ground-based images covering a wide field of view (to select the brightest objects at $z$$\sim$6) and deeper {\it Hubble}/WFC3 and ACS data with a smaller field of view (to detect fainter objects). 

In Section~\ref{obs}, we present our new data used for the analysis and their reduction procedure.
In Section~\ref{LBG}, we explain in detail the method that was used to select $z$$\sim$6 objects and we present the candidates coming from the two ground and space-based
datasets. Their photometric properties are presented in Section~\ref{sedfitting}. In Section~\ref{environment} we discuss the presence of an over-density of LBGs
in the \textit{HFLS3} field. The concordance cosmology is adopted throughout this paper, with $\Omega_{\Lambda}$=0.7, $\Omega_{m}$=0.3 and H$_0$=70 km s$^{-1}$ Mpc$^{-1}$. All magnitudes are quoted in the AB system \citep{AB}.

\section{Observations and data reduction}
\label{obs}
In this Section, we present the ground-based and space-based data we used to perform the search for $z$$\sim$6 objects around \textit{HFLS3}. GTC optical and Spitzer data have already been presented in \citet{Riechers13}. We have re-reduced the GTC data to improve their image qualities and we include a newly acquired K$_s$-band image
taken with LIRIS on the the 4.2m William Herschel Telescope (WHT).
In the following we summarize image properties and the data reduction procedure. A summary of the data used in the current study
is listed on Table~\ref{photometric_properties}.

\subsection{Ground-based data}
In order to select bright objects at $z$$\sim$6 and to establish the bright-end of the
UV luminosity function \citep{Lacey11}, we performed a first search for LBGs in ground-based images taken with OSIRIS on GTC (\citealt{GTC}, \citealt{OSIRIS}) and  LIRIS installed on WHT (\citealt{WHT}, \citealt{LIRIS}). 

OSIRIS data were acquired between June 29 and  August 3 2011, as part of the GTC2-10ITP program (P.I.: I. P\'erez-Fournon), and used $g'$, $r'$, $i'$and $z'$ broad-band filters. The field of view covered by OSIRIS is 7.8' x 7.8' and the total exposure times in each band are 2.7 ksec in $g'$ and $r'$, 21.6 ksec in $i'$ and 18,7 ksec in $z'$. Each frame was reduced individually following standard reduction procedure  in IRAF\footnote{http://iraf.noao.edu/} (bias subtraction, flat fielding, sky subtraction) and registration and combination using SCAMP  \citep{SCAMP} and SWARP \citep{SWARP}. To reduce the sky background at long wavelengths we applied for the $i'$ and $z'$ band images, a double sky subtraction using the \textit{IRAF} package \object{XDIMSUM}, with steps explained in \citet{Richard06}. The astrometric and photometric calibrations were performed using the 9$^{th}$ release of the SDSS catalog \citep{SDSS}. Moreover to produce the best $z-$ band quality image, we applied the "best seeing stacks" method described in \citet{CFHTLS}, selecting only the best frames in terms of seeing. We measured the FWHM on each frame using PSFEx \citep{psfex}, and we kept 86.4\% of our original dataset beyond which the FWHM increase without a significant evolution of the image depth. The seeing of the final images is 0.89'', 0.83'', 0.83'' and 0.89'' respectively in $g'$, $r'$, $i'$ and $z'$. The depth was computed using empty 1.4'' radius apertures all over the field. 

We added to our previous dataset a new near-IR, K$_s$ image acquired on June 5 2014 with the LIRIS instrument  (4.27'$\times$4.27' field of view, 0.25'' size pixels) installed at the Cassegrain focus of the WHT, as part of a new WHT Large Program on HerMES high-redshift galaxies (P.I.: I. P\'erez-Fournon). The total integration time was 3.6 ksec, the observing conditions were photometric and the seeing was 0.5". The reduction of the image was performed using the IAC's IRAF lirisdr task\footnote{www.iac.es/galeria/jap/lirisdr/LIRIS$\_$DATA$\_$REDUCTION.html} and the astrometry calibration was carried out using the Graphical Astronomy and Image Analysis Tool (GAIA) included in the Starlink astronomical software package\footnote{http://star-www.dur.ac.uk/~pdraper/gaia/gaia.html}$^,$\footnote{http://starlink.jach.hawaii.edu/starlink}.  The K$_s$ image was matched to SDSS DR10 stars and is estimated to be accurate to 0.1" (rms of the fit). The photometric calibration of the image was done with 2MASS stars. The 5-sigma depth of this image in 1.4" radius apertures is 23.14.

\subsection{Space-based data}

The data reduction of the ACS and WFC3 data we used in this study are described in detail in \citet{Cooray14}. We summarize hereafter the principal steps. These data are part of an HST Cycle 21 program (GO 13405, PI: A. Cooray) and used 6 orbits to produce images of the \textit{HFLS3} environment in F625W, F814W, F105W, F125W and F160W (Table \ref{photometric_properties}). We used the IRAF.STSDAS pipeline and the CALWFC3 tool for flat-fielding and cosmic-rays rejections and ASTRODRIZZLE \citep{Fruchter10} to combine individual exposure and create 0.06''/pixel WFC3 images. The ACS data were flat-fielded, charge transfer efficiency corrected and cosmic rays rejected using the CALACS pipeline (version 2012.2), and then combined with ASTRODRIZZLE to produce 0.03''/pixel images. Astrometric calibration was performed using SDSS (9$^{th}$ release). The WFC3 and ACS data were matched independently following the classical IRAF procedure. The depth was computed using empty apertures all over the field (0.25'' radius for ACS and 0.50'' radius for WFC3 images). 

In order to extend the wavelength coverage of our survey, we also used data obtained with IRAC \citep{Fazio04} on board the \textit{Spitzer Space Telescope} as part of 
a DDT program (ID: 80240 - PI: J. Vieira) on 2012 March 21. The on-target observations consisted of 38 frames with an integration time of 100 sec 
in each of the 2 channels centered at 3.6 $\mu$m (ch1) and 4.5 $\mu$m (ch2). We used corrected  Basic Calibrated Data (cBCD) frames which are already corrected by pipeline for various artifacts, such as multiplexer bleed and pulldown. These images, together with associated mask and uncertainty images, were processed, drizzled (with a PIXFRAC of 0.65), and combined using the standard pipeline MOPEX. The final mosaics images have a pixel size of 0.6", roughly a half of the IRAC native pixel scale. The mosaic has a 3$\sigma$ point source sensitivity of 0.384 $\mu$Jy and  0.412 $\mu$Jy in 3.6$\mu$m and 4.5$\mu$m respectively.

\begin{figure}
\centering
\includegraphics[angle=0,scale=.65,width=8.9cm]{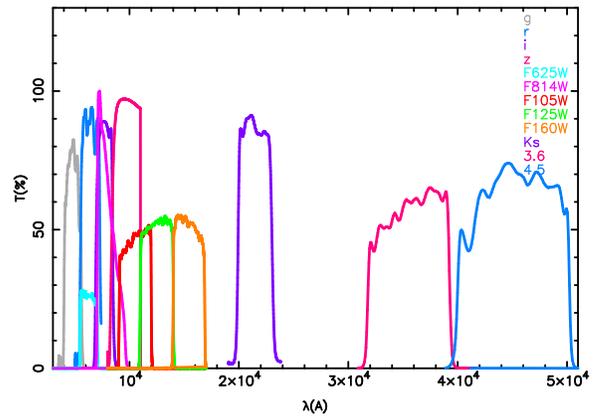}
\caption{ \label{transmission} Wavelength coverage and filters transmission.}
\end{figure}

\begin{deluxetable}{cccccc}
\tabletypesize{\scriptsize}
\tablecaption{Photometric properties of the imaging data}
\tablewidth{0pt}
\tablehead{
\colhead{Filter} & \colhead{$\lambda_{eff}$} & \colhead{$\Delta\lambda$} & \colhead{t$_{exp}$}  &
\colhead{m(5$\sigma$)} & \colhead{Instrument}  \\
\colhead{} & \colhead{[nm]} & \colhead{ [nm]} & \colhead{[ks]} &\colhead{} & \colhead{}\\
\colhead{(1)} &\colhead{(2)} &\colhead{(3)} &\colhead{(4)} &\colhead{(5)} &\colhead{(6)} 
}

\startdata
$g$'	&	481.5	& 153	& 2.7			& 26.91$^a$	&	OSIRIS/GTC \\
$r$'	&	641.0	& 176	& 2.7			& 26.80$^a$	&	OSIRIS/GTC \\
$i$'	&	770.5	& 151	& 21.6		& 26.60$^a$	&	OSIRIS/GTC \\
$z$'	&	969.5	& 261	& 16.3		& 25.87$^a$	&	OSIRIS/GTC \\ \hline
$F625W$	&	629.6	& 98	& 1.6			& 26.08$^b$	&	ACS/HST \\ 
$F814W$	&	811.5	& 166& 	2.3		& 26.99 $^b$	&	ACS/HST\\ 
$F105W$	&	1055.1	& 265& 	9.9		& 25.90$^c$	&	WFC3/HST \\ 
$F125W$	&	1248.6	& 443& 	4.4		& 26.30$^c$	&	WFC3/HST \\ 
$F160W$	&	1536.9	& 268& 	2.8		& 26.00$^c$	&	WFC3/HST \\  \hline
$Ks$		& 	2150.0	& 320& 	3.6		& 23.14$^a$	&	LIRIS/WHT \\  \hline
3.6$\mu$m	&3575.0	& 776& 	3.8		& 24.38$^d$	&	IRAC/Spitzer \\ 
4.5$\mu$m	&4528.0	& 1060& 	3.8		& 24.30$^d$	&	IRAC/Spitzer \\ \hline

\enddata
\tablecomments{\label{photometric_properties} (1) filter identification, (2) filter central wavelength, (3) filter width, (4) exposure time, (5) 5$\sigma$ AB magnitude and (6) instrument and telescope. }
\tablenotetext{a}{point-source depth measured in a 1.4'' radius aperture}
\tablenotetext{b}{point-source depth measured in a 0.25'' radius aperture}
\tablenotetext{c}{point-source depth measured in a 0.50'' radius aperture}
\tablenotetext{d}{point-source aperture corrected depth measured in a 1.9'' radius aperture}
\end{deluxetable}

\subsection{Source catalogs}
\label{extraction}

We used \textit{SExtractor} 2.18.4 \citep{sextractor} to build catalogs in double image mode using a $\chi^2$ image including all individual picture, as detection picture \citep{Szalay99}. The extraction parameters we used have been chosen in order to extract faint objects. Therefore the DETECT\_MINAREA parameter was fixed to 4.0 pixels according to the seeing of our images, and to limit spurious detections we choose DETECT\_THRESH =1.7$\sigma$ (e.g. see \citealt{Muzzin13}) DEBLEND\_NTHRESH=16 and DEBLEND\_MINCONT=0.00002 \citep{Scoville07}. We measured the photometry in apertures with radius defined by 2$\times$FWHM on PSF-matched data on the GTC data and 0.4'' radius aperture on HST images, as well as with \textit{SExtractor}  MAG\_AUTO. We found that these two types of magnitude are in good agreement for all point-source like objects in our catalogs.

\section{Selection of $z$$\sim$6 Lyman Break Galaxies}
\label{LBG}
The LBG technique \citep{LBG} has been extensively used to select galaxies at high-$z$ (e.g. \citealt{Castellano10},  \citealt{Trenti12}, \citealt{Laporte14}).
It combines two different criteria: non-detection/detection and the color selection. We applied this method on \textit{SExtractor} catalogs \citep{sextractor} built using detection parameters as a function of the dataset we used (see below for details). Regarding the depth and spatial resolution of each dataset, we performed independent searches in each survey: from our ground-based images we focussed on the brightest objects and from the HST data we investigated faintest sources at the redshift of \textit{HFLS3}. 

\subsection{Selection from \textit{GTC} data}
\label{color}
The GTC/OSIRIS imaging data we used to search for LBG at the redshift of \object{HFLS3} covered a wavelength range between 328.5 and 1230.5 nm. The non-detection/detection criteria we applied to our catalogs to select $z$$\sim$6 objects were the following:
\begin{equation}
\centering
m_{g'}>m(2\sigma)_{g'} = 27.9
\end{equation}
\begin{equation}
\centering
m_{r'}>m(2\sigma)_{r'} = 27.8
\end{equation}
\begin{equation}
\centering
m_{z'}<m(5\sigma)_{z'} = 25.9
\end{equation}
The lack of deep near-IR data over the OSIRIS field-of-view does not allow us to produce a color-color diagram with a clear selection window. 

\begin{figure}
\includegraphics[angle=0,scale=.65,width=8.2cm]{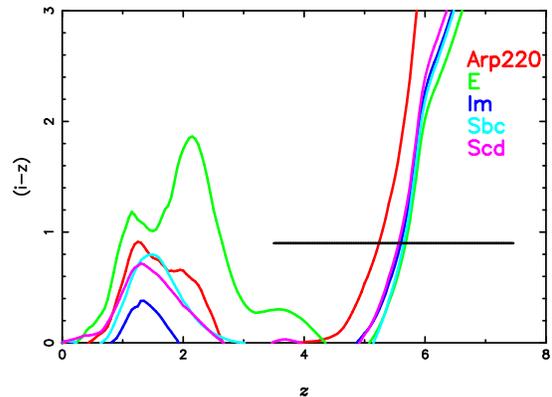}
\caption{\label{color_GTC} $i$'-$z$' color cut defined for the $z$$\sim$6 LBG selection using GTC filters and color evolution of several templates of galaxies (\citealt{BC} and \citealt{Polletta07}). The color-criteria is defined by the black line, and shows that in that case the majority of the contaminants are elliptical galaxies at mid-$z$s.}
\end{figure}

\begin{figure}
\includegraphics[angle=0,scale=.65,width=8.2cm]{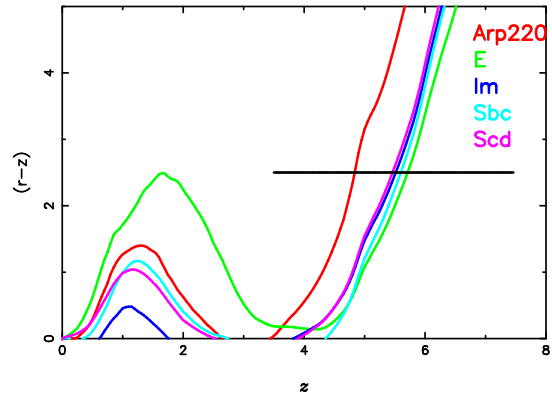}
\caption{\label{color_GTC_rz} $r$'-$z$' color window defined for the $z$$\sim$6 LBG selection using GTC filters.}
\end{figure}

\begin{figure}
\includegraphics[angle=0,scale=.35]{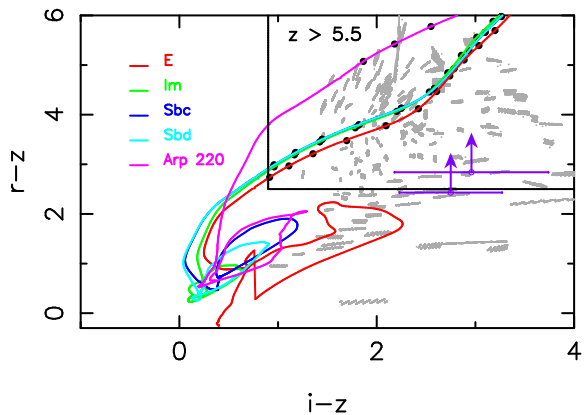}
\caption{\label{rziz} Color criteria used to select $z>$5.5 objects using templates from \citet{BC} and \citet{Polletta07}. Position of the two GTC candidates are displayed by magenta dots. We over-plotted the expected color of M, L and T dwarfs in grey computed from 225 stellar spectra (see references in Sec. \ref{contaminants}). Black dots show the colors of objects with redshift ranging from 5.5 to 6.5 per bin of 0.1  }
\end{figure}

We defined a color window using the $i'-z'$ color evolution of a set of templates (\citealt{BC} and \citealt{Polletta07}) as a function of the redshift (figure\ref{color_GTC};  m$_i$-m$_z > $0.9). These criteria limit the contamination by most of the low-$z$ interlopers except for  the elliptical galaxies at $z<$3. In order to avoid the contamination by such interlopers, we estimated the size of the break between optical and NIR beyond which the selection of mid-$z$ interlopers is unlikely  (figure \ref{color_GTC_rz}). The break required by the evolution of templates mentioned above (m$_r$-m$_z>$2.5 mag) is consistent with the non-detection criteria that we applied, and thus by combining the previous $i'-z'$ color criterion (m$_i$-m$_z>$0.9 mag) with the non-detection/detection criteria listed above, we eliminate a large number of mid-$z$ interlopers. Therefore, we can summarize the color selection we defined to select $z$$\sim$6 objects on the GTC data as follows: \\
\indent m$_i$ - m$_z$ $>$ 0.9 \\
\indent m$_r$ - m$_z$ $>$ 2.5 \\
The selection window is shown on Figure \ref{rziz}.

\begin{figure*}
\centering
\includegraphics[angle=0,scale=.65,width=15cm]{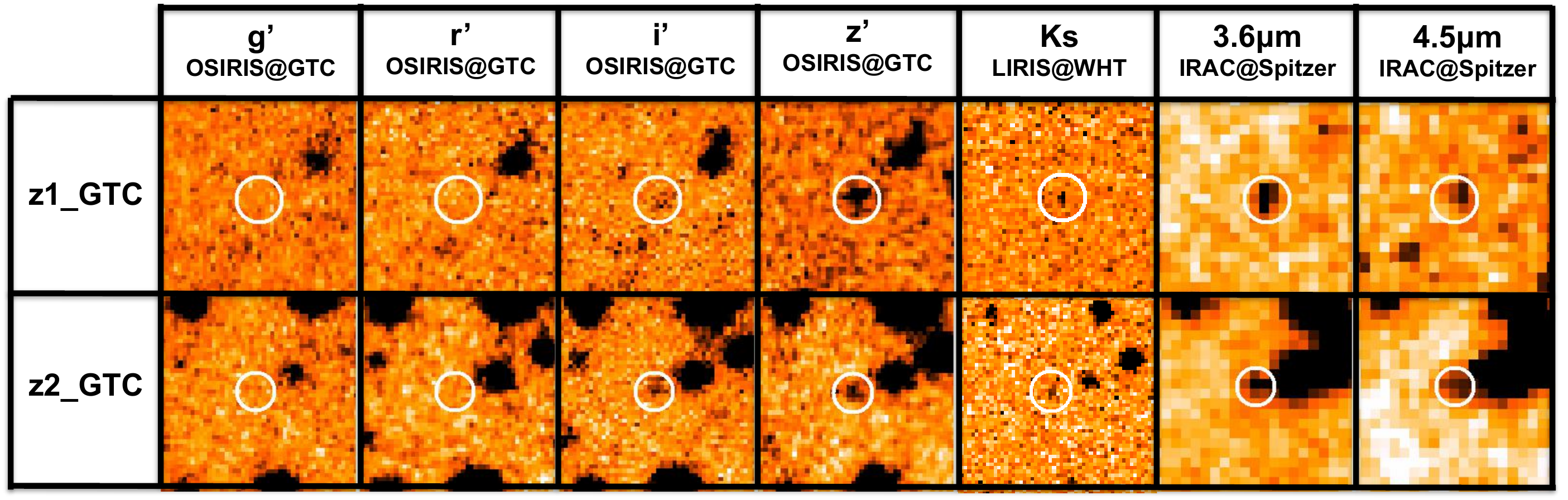}
\caption{\label{gtc_trombi} Thumbnail images of the two candidates selected in the 7.8'$\times$7.8' OSIRIS field of view. The size of each stamps is 12''$\times$12''. The position of the candidate is displayed by a white circle of 1.4'' radius aperture.}
\end{figure*}

After visual inspection only two objects satisfy the selection criteria defined above with $m_z<$25.1. These two candidates are displayed as z1\_GTC and z2\_GTC in Figure \ref{gtc_trombi} and Table \ref{gtc_drops_table}.

\begin{deluxetable*}{cccccccccc}
\tabletypesize{\scriptsize}
\tablecaption{Photometry of the two candidates selected from the GTC/OSIRIS data}
\tablewidth{0pt}
\tablehead{
\colhead{ID} & \colhead{RA} & \colhead{DEC} & \colhead{m$_g$} & \colhead{m$_r$} & \colhead{m$_i$} & \colhead{m$_z$} & \colhead{m$_{Ks}$} & \colhead{m$_{3.6}$} & \colhead{m$_{4.5}$}  \\
}
\startdata
z1\_GTC	& 17:06:40.6	& +58:47:49.6 & $>$27.9	&	$>$27.8	&	27.92	& 24.96		& 23.08	& 23.24		& 24.73 \\
		&	&&		&			&  $\pm$0.77	& $\pm$0.10	& $\pm$0.22	& $\pm$0.08	& $\pm$0.22 \\
z2\_GTC	& 17:06:41.310 	& +58:47:17.28 &$>$27.9	&	$>$27.8	&	27.48	& 25.05		& 23.62	& 22.91	& 23.00 \\
		&	&&		&			&  $\pm$0.51	& $\pm$0.11	&$\pm$0.36	& $\pm$0.20 	& $\pm$0.20 \\
\enddata
\tablecomments{\label{gtc_drops_table} All GTC-magnitudes are MAG\_AUTO magnitude from \textit{SExtractor}. Upper limits are 2$\sigma$ depth. Error bars are computed within empty 1.4'' radius apertures around the object on the GTC data. IRAC magnitudes are measured in a 1.9'' radius aperture and error bars are computed using empty 1.9'' radius aperture around the object. }
\end{deluxetable*}


\subsection{Selection from \textit{HST} data}

\begin{figure}
\centering
\includegraphics[angle=0,scale=.65, width=9.5cm]{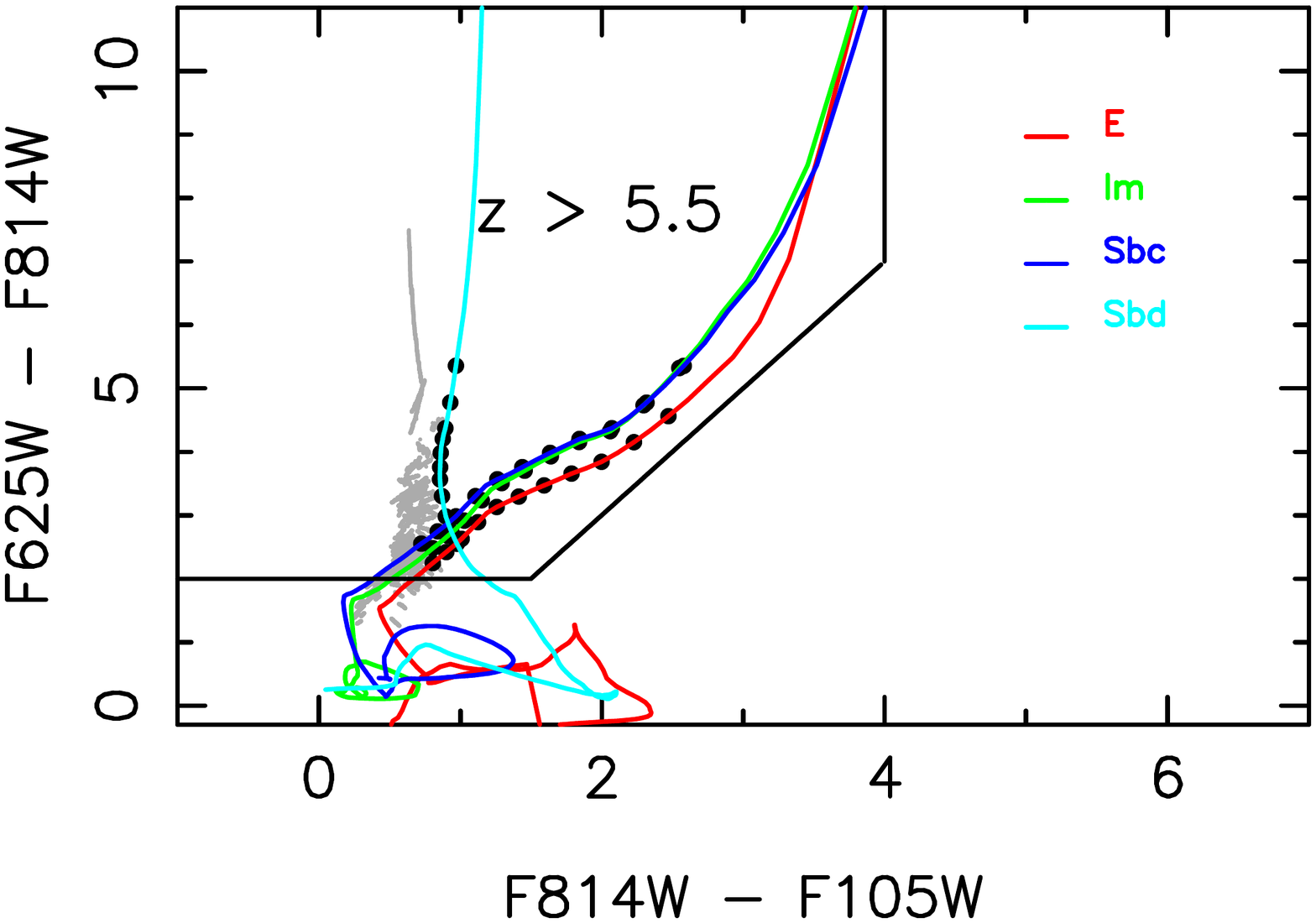}
\caption{\label{color_HST_riY} F625W-F814W-F105W color window defined for the $z$$\sim$6 LBG selection using HST filters and color evolution of several templates of galaxies given by \citet{BC} and \citet{Polletta07}. The color-criteria for $z>$5.5 are defined by the black box. We over-plotted the expected color of M, L and T dwarfs in grey computed from 225 stellar spectra (see references in Sec. \ref{contaminants}). Black dots show the colors of objects with redshift ranging from 5.5 to 6.5 per bin of 0.1 } 
\end{figure}

The second sample has been built using \textit{HST} data covering wavelength from 580.6 to 1670.9 nm. In order to select $z$$\sim$6 objects and regarding the set of filters we have, we defined the following detection/non-detection criteria:
\begin{equation}
\centering
m_{F625W}>m(2\sigma)_{F625W} \text{  } \cup \text{  }
m_{F105W}<m(5\sigma)_{F105W}
\label{det_HST}
\end{equation}
or in terms of color :
\begin{equation}
\centering
m_{F625W} - m_{F105W} > 1.2 \,.
\end{equation}
The wavelength coverage of the \textit{HST} data is better in the near-IR domain compared to the \textit{GTC} survey. Therefore we can define a color-window to select $z$$\sim$6 objects following the standard method described in section \ref{color}. Figure \ref{color_HST_riY} displays this selection window defined by:
\begin{align}
m_{F625W}-m_{F814W} &>2.0 \\
m_{F814W}-m_{F105W} &<4.0 \\
m_{F625W}-m_{F814W} &>0.57\times (m_{F814W}-m_{F105W}) + 0.66
\end{align}

The use of broad-band filters as well as color selection imply that several of the $z$$\sim$6 candidates will be lost during the selection process. In order to estimate the fraction of those objects, we computed the incompleteness of the selection function described in this paper. Using templates from Coleman et al. (1980), Kinney et al. (1996), Poletta el al. (2007) and Silva et al. (1998) and transmission of filters used in this survey, we simulated 100 000 objects in the redshift range 4.0 $< z <$ 8.0 per bin of 0.25 magnitude with m$_{F105W}$ ranging from 19.0 to 29.0. The selection by the color-criteria defined above shows that $\sim$74\% of objects at $z$$\sim$6.3 are selected. 

As previously we used \textit{SExtractor} 2.18.4 in double image mode using a $\chi^2$ image \citep{Szalay99} made with the WFC3 images as detection picture to produce WFC3 catalogs, because $z$$\sim$6 LBGs should be detected at these wavelengths. As in the "ground-based" selection, we used \textit{SExtractor} parameters defined to maximize the selection of faint objects. We then used TOPCAT \citep{Taylor05} to match ACS single image mode catalogs with the WFC3 catalogs allowing a search radius fixed to 2$\times$FWHM of the worst seeing image. The total magnitude were obtained from aperture correction using F160W MAG\_AUTO and following the method described in \citet{Finkelstein13}.   Each catalog included $\sim$1600 detections. Colors were measured on psf-matched data, and 14 objects follow the criteria defined above. Among these sources, one of our GTC candidates is covered by the HST field of view and is included in this new sample (ID: z2$\_$GTC), showing that our two selection functions are well defined to select this kind of object. The other GTC candidate (z1$\_$GTC) is not detected on the F625W images and well detected on the F814W (m$_{F814W}$= 26.80$\pm$0.18) confirming the break between the $i'$ and $z'$-band of our GTC survey. However, \textit{HFLS3} is not included in this sample, because \textit{SExtractor} failed to extract it properly, it is blended with the 2 nearby $z\sim$2.1 galaxies acting as a foreground lens (see \citealt{Cooray14} for more details). Five objects have been removed from this sample because they are detected on the deep $g'$ band data from the GTC survey (see section \ref{contaminants} for details). Therefore the final sample is composed of nine objects and is presented on Table \ref{hst_drops_table} and Figure \ref{trombi_hst}.

\subsection{Detection in \textit{LIRIS} and \textit{IRAC} data} 

We increased the number of SED constraints in the near-IR domain by using data from \textit{LIRIS} and \textit{IRAC}. Due to the depth of our K$_s$ image, only the two GTC-objects are clearly detected in the near-IR (including the candidate in common in the two samples). These results are not surprising regarding the brightness of these sources, but it could have been a way to remove mid-$z$ interlopers from the HST sample. Indeed previous papers have shown that mid-$z$ interlopers display a very high-$z$ LBG like SED up to 1.6$\mu$m, but they are very red at larger wavelengths corresponding to IRAC (e.g. \citealt{Boone11})  The depth of IRAC data (m$_{3.6, 4.5}$(5$\sigma$)$\sim$24.3) is not completely sufficient to add robust constraints on the SEDs of all our candidates. Among the two different samples, two objects are clearly detected in the IRAC data. We extracted the photometry of z2\_GTC using \textit{imfit}\footnote{www.mpe.mpg.de/~erwin/code/imfit/index.html}. For both  objects, we measured the photometry in a 1.9'' radius aperture, and used the correction factor computed by \citet{Mauduit12}. The error bar was computed using empty 1.9'' radius apertures around the object. 

\subsection{Contaminants}
\label{contaminants}

The most common sources of contamination of high-$z$ samples are SNe, AGN, low-mass stars, photometric scatter, transient objects, spurious sources, extremely red galaxies, among others. All our candidates are detected in at least two bands limiting the contaminations by spurious sources, moreover these observations spread over several months limiting the detection of transient sources. However, we noted that our ample is not free of AGN and SN contamination given the timescale during which our observations have been carried out. Regarding the field of view covered by these two surveys and the statistical number of supernovae expected in $\sim$1deg$^2$, the probability of being contaminated by a supernova is relatively small. Low-mass stars, such as M, L and T dwarfs, could display the same colors properties as those expected for high-$z$ galaxies. We computed the expected colors of these galaxies through all the filters we used in this study from 225 stellar spectra ( \citealt{Burgasser06b} , \citealt{Burgasser04}, \citealt{Burgasser08}, \citealt{Burgasser07}, \citealt{Burgasser06a}, \citealt{Cruz04}, \citealt{Kirkpatrick10}, \citealt{Reid06}, \citealt{Siegler07}, \citealt{Chiu06}, \citealt{Looper07}, \citealt{McElwain06}, \citealt{Sheppard09}, \citealt{Liebert07}, \citealt{Burgasser06}). Fig. \ref{color_HST_riY} and \ref{rziz} display the expected colors of these stars that are consistent with previous studies (e.g. \citealt{Finkelstein14}, \citealt{Willott13}) and show that stellar contamination is non-negligible is both samples. Therefore to limit the selection of low-mass stars, expected to be unresolved on our dataset, we used SExtractor FLUX\_RADIUS enclosing 50\% of the total flux and the SExtractor stellarity parameter. Both indicators demonstrate that two objects among our samples are likely unresolved, namely z1\_HST and z3\_HST with a stellarity parameter of $>0.7$ and a size computed from the half light-radius comparable with the FWHM of the image, taking uncertainties into account. 

Therefore the majority of contaminants that can enter into our sample involves the extremely red mid-$z$ galaxies. We estimated the contamination rate of our sample by mid-$z$ interlopers by using the colors distribution of mid-$z$ objects over the fields of view covered by our two datasets (e.g. \citealt{Oesch10}). We followed a 3 steps method summarized hereafter : (1) we selected all objects that are detected at more than 2$\sigma$ in all bands, (2) then we matched the luminosity range to the luminosity covered by our two samples and added the corresponding uncertainties on the photometry, (3) finally we applied the selection criteria we used to build our high-$z$ candidates samples and all the selected objects are mid-$z$ interlopers. 3 objects are identified over the 4.5 arcmin$^2$ covered by HST data and 2 using our GTC data leading to $\sim$ 7-36\% and $>$20\% contamination rate including cosmic variance respectively. It has been shown by several recent studies (e.g. \citealt{Hayes12}, \citealt{Laporte15}) that the best way to remove such objects is to cover the short wavelength with deeper imaging data, regarding that the contamination rate by such extreme mid-$z$ interlopers is rather uncertain. We used our deep GTC $g'$-band data to confirm the real non-detection of HST candidates. We find that one of our targets is clearly detected (m$_{g}\sim$4$\sigma$) and can not, therefore, be at $z$$\sim$6. Four others candidates are faint on that image (m$_{g'}\sim$2$\sigma$) and thus cannot also be at such high-$z$, but further data are necessary to confirm their optical emission.

\begin{figure*}
\centering
\includegraphics[angle=0,scale=.45]{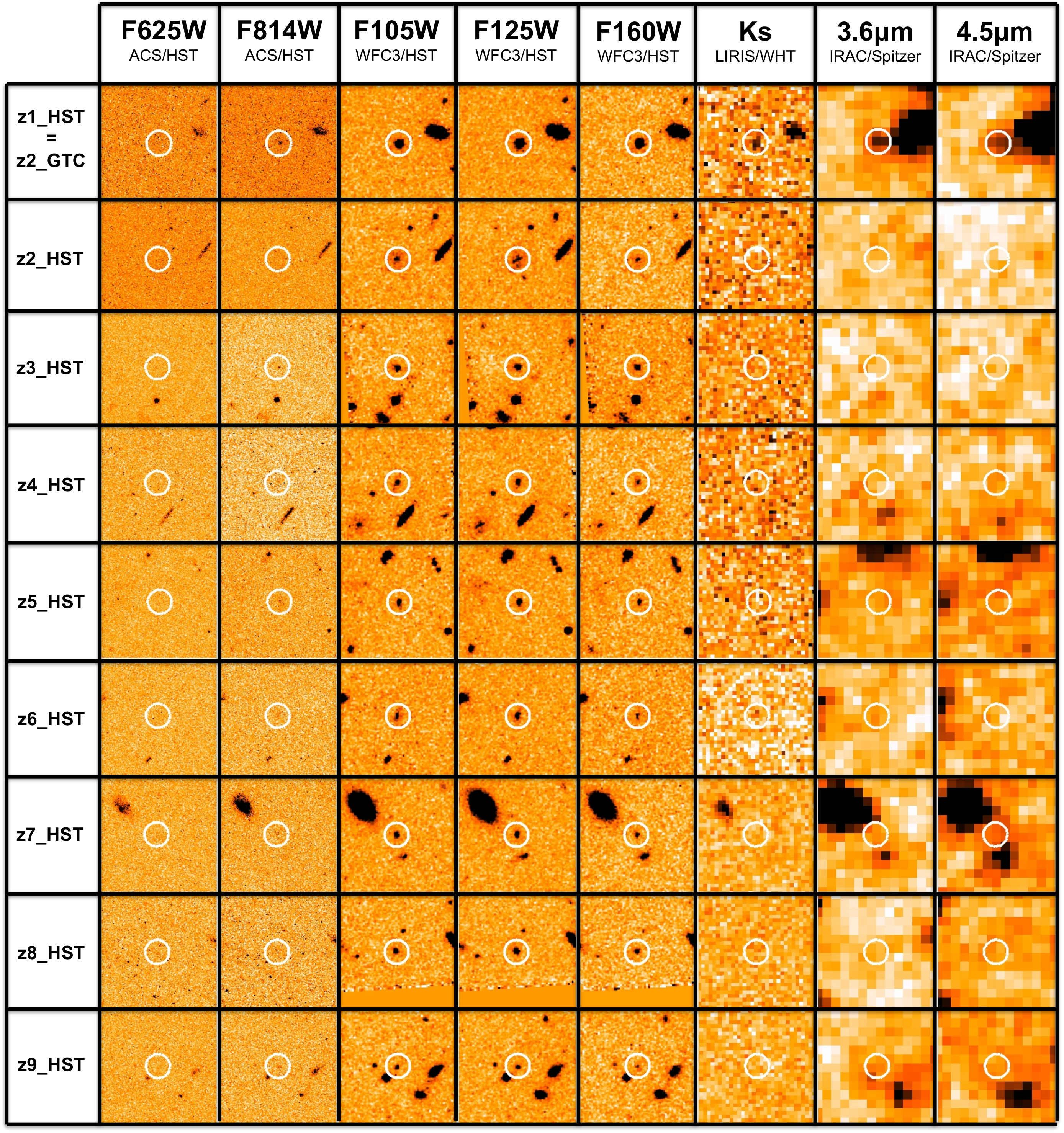}
\caption{\label{trombi_hst} Thumbnail stamps image of the nine candidates selected in the WFC3 field of view. The size the stamps are 7.5''$\times$7.5'' in all the bands. The position of the candidate is displayed by a white-circle of 0.8'' radius aperture . }
\end{figure*}

\begin{figure*}
\centering
\includegraphics[angle=0,scale=.65]{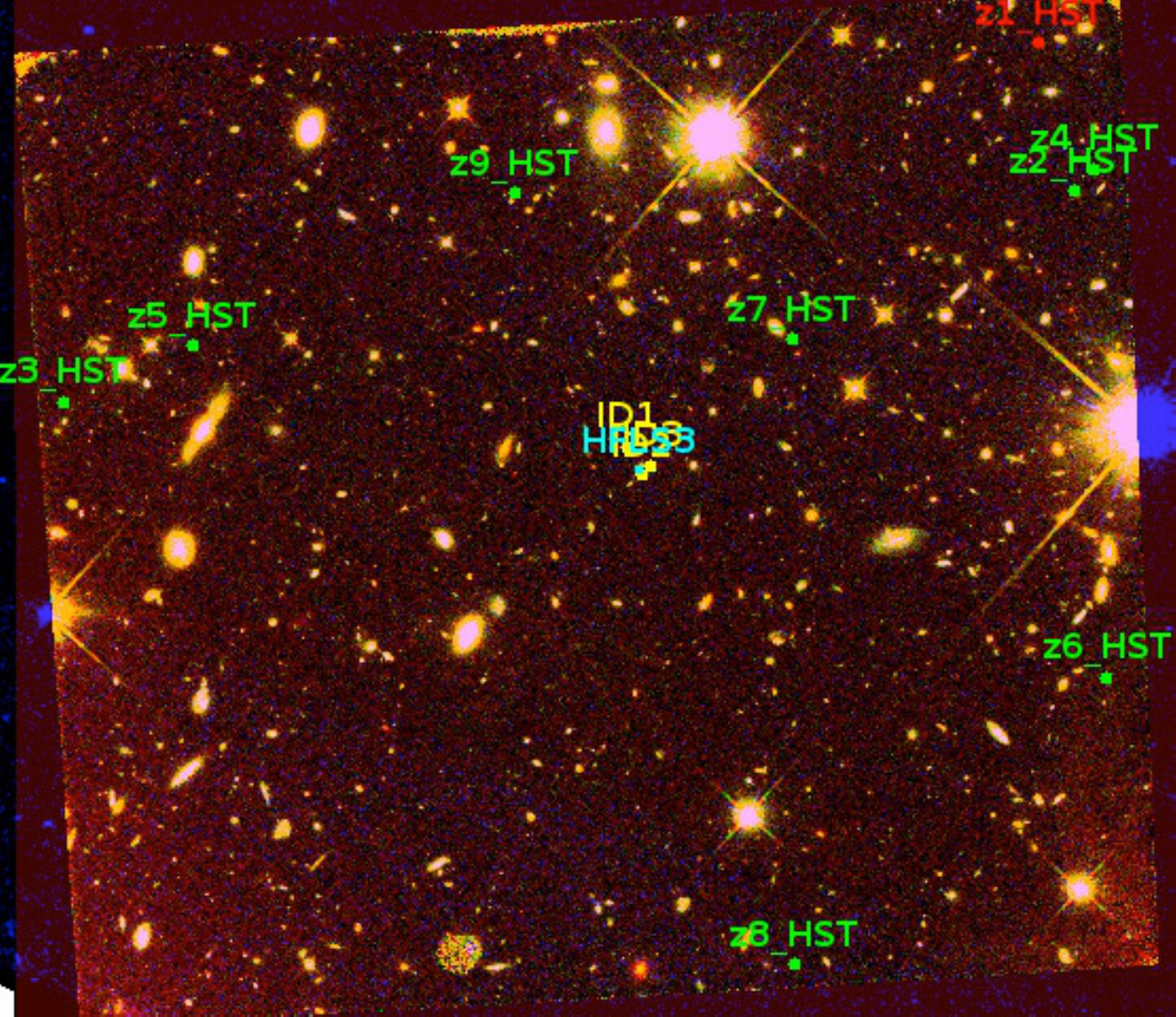}
\caption{\label{distribution} Distribution of our candidates around the sub-millimeter starburst \textit{HFLS3} on a composit image showing the field-of-view 
covered by our HST/WFC3 data. The position of each candidate is displayed by a green circle (red for the candidate in common in GTC and HST samples) and the \textit{Herschel} starburst by a cyan circle. We also plotted the position of the three faint objects discussed in \ref{faint}}
\end{figure*}

After removing these mid-$z$ interlopers, the final HST sample is composed of nine objects with m$_{F105}$ ranging from 24.5 to 26.5 (figure \ref{distribution}). The postage stamps are displayed on Figure \ref{trombi_hst} and Table \ref{hst_drops_table} presents the photometry of this sample.

We note here that our ACS images are not deep enough to exclude definitively the low-$z$ hypothesis for all our $z$$\sim$6 galaxy candidates. Indeed, the 2$\sigma$ depth of the F625W image used here (27.1 AB) combined with the F814W depth (27.99 AB) does not allow us to apply completely the color-color criteria we imposed (F625W - F814W $>$2.0). For most of them, the F625W-F814W break is not high enough and even if we used the 1$\sigma$ detection as upper limit in F625W we never reach a break of 2 magnitudes in these two filters (as in previous studies, e.g. \citealt{Monna14}). This is why we primarily use here the deeper ground-based GTC $g$-band data to identify the
interlopers.

\begin{deluxetable*}{ccccccccccc}

\tabletypesize{\scriptsize}
\tablecaption{Photometry of the candidates selected in the HST data}
\tablewidth{0pt}
\tablehead{
\colhead{ID} & \colhead{RA} & \colhead{DEC} & \colhead{m$_{F625}$} & \colhead{m$_{F814}$} & \colhead{m$_{F105}$} & \colhead{m$_{F125}$} & \colhead{m$_{F160}$} & \colhead{m$_{3.6}$} & \colhead{m$_{4.5}$}  
}
\startdata
z1\_HST	& 17:06:41.310 	& +58:47:17.28 	& $>$27.08 	& 26.42		& 24.52 		& 24.09 		& 23.90 		& 22.91		& 23.00  		\\ 
	&		&							&			& $\pm$ 0.13	& $\pm$ 0.07	&$\pm$ 0.03	&$\pm$ 0.03	&$\pm$0.20     & $\pm$0.20		\\
z2\_HST	& 17:06:40.737 	& +58:46:58.52 	& $>$27.08 	&  $>$27.99	&25.08		&25.37		&25.08		&$>$25.38	&$>$25.30	\\
	&		&		&								& 			& $\pm$ 0.11	&$\pm$ 0.10	&$\pm$ 0.09	&		& 		\\
z3\_HST	& 17:06:57.157 	& +58:46:31.85 	& $>$27.08 	& 27.01		&25.34		&25.07		&25.09		&$>$25.38	&$>$25.30	\\
	&		&		&								& $\pm$ 0.23	& $\pm$ 0.13	&$\pm$ 0.07	&$\pm$ 0.09	&		& 		\\
z4\_HST	& 17:06:40.411 		& +58:47:01.39 	& $>$27.08 	& 27.91		&25.81		&25.53		&25.53		&$>$25.38	&$>$25.30	\\
	&		&		&								& $\pm$ 0.47	& $\pm$ 0.20	&$\pm$ 0.11	&$\pm$ 0.14	&		& 		\\
z5\_HST	& 17:06:55.049 	& +58:46:39.05 	& $>$27.08 	& $>$27.99	&25.87		&25.61		&25.60		&$>$25.38	&$>$25.30	\\
	&		&		&								&			& $\pm$ 0.21	&$\pm$ 0.12	&$\pm$ 0.15	&		& 		\\
z6\_HST	& 17:06:40.212 	& +58:45:57.16 	& $>$27.08 	& $>$27.99	&25.92		&25.64		&25.81		&$>$25.38	&$>$25.30	\\
	&		&		&		& 									& $\pm$ 0.23	&$\pm$ 0.12	&$\pm$ 0.18	&		& 		\\
z7\_HST	& 17:06:45.315 	& +58:46:39.78 	& $>$27.08 	& 26.73		&25.94		&25.67		&25.68		&$>$25.38	&$>$25.30	\\
	&		&		&		& 						$\pm$ 0.15	& $\pm$ 0.23	&$\pm$ 0.13	&$\pm$ 0.16	&		& 		\\
z8\_HST	& 17:06:45.289 	& +58:45:20.86 	& $>$27.08 	& 27.67		&26.11		&25.80		&26.01		&$>$25.38	&$>$25.30	\\
	&		&		&								& $\pm$ 0.37	& $\pm$ 0.27	&$\pm$ 0.14	&$\pm$ 0.22	&		& 		\\
z9\_HST	& 17:06:49.819 	& +58:46:58.26 	& $>$27.08 	&26.86	 	&26.47		&26.45		& 26.43	&$>$25.38	&$>$25.30	\\
	&		&		&								& $\pm$ 0.18	& $\pm$ 0.37	&$\pm$ 0.25	& $\pm$0.32		&		& 		\\

\enddata
\tablecomments{\label{hst_drops_table} All HST-magnitudes are MAG\_AUTO magnitude from \textit{SExtractor}. Upper limits are 2$\sigma$ depth. Error bars are computed within empty 0,125'' radius apertures around the object on the ACS data and 0.25'' radius on WFC3 data. IRAC magnitudes are measured in a 1.9'' radius aperture and error bars are computed using empty 1.9'' radius aperture around the object. }
\end{deluxetable*}

\subsection{Expected number of objects}

The effective surface covered by our OSIRIS survey is 55.5 arcmin$^2$  and has been computed by masking all the bright objects on our $\chi^2$ detection picture. We computed the expected number of sources in that field of view by using the shape of the luminosity function published by \citet{Bouwens12}, which is relatively well established at $z$$\sim$6. Including cosmic variance and uncertainties on the Schechter parameters, the expected number of objects in the range of redshift 5.5$< z <$6.5 and with m$_z<$25.9, is 1.3$^{+2.0}_{-0.9}$, showing that our sample built from GTC data is in excellent agreement with the expectation. For the HST-survey covering an effective surface of 4.54 arcmin$^2$, the expected number of objects detected at most than 5$\sigma$ with WFC3 bands in the same area is 6.10$^{+9.0}_{-4.1}$ (including cosmic variance). Therefore our selection is in relatively good agreement with the expectations, without any evidence for a significant overdensity. 


\section{Photometric properties of the $z$$\sim$6 candidates}
\label{sedfitting}

This Section presents the principal photometric properties of the selected objects using a SED-fitting approach. We discuss in the following the photometric redshift, 
UV luminosity, SFR, and the reddening for our two samples.


We used the Version 12.2 of \textit{Hyperz}\footnote{http://userpages.irap.omp.eu/~rpello/newhyperz/} \citep{Hyperz} with the standard templates library including Bruzual \& Charlot (2003), Coleman et al. (1980), Kinney et al. (1996), Poletta el al. (2007) and Silva et al. (1998) templates. The redshift space parameters ranges from $z\sim$0 to 8 and $A_v$ from 0.0 to 3.0 mag. The best-fit is always found at $z>$5.3 with a reasonable reddening solution for most of the candidates (A$_v<$1.00mag). The redshift probability distribution, hereafter P(z), is well defined around the best photo-$z$ and no low-$z$ solution appears clearly for all our targets. For the candidate in common in the two samples, we estimated its properties in three cases: (1) combining all of the data points we have, (2) using only the GTC, WHT and Spitzer photometry, and (3) using the HST, WHT and Spitzer constraints. The results are similar and give a moderate reddening solution (A$_v\sim$1.80-2.0) and a best photo-$z$ included between 5.8 and 6.1. 

As a cross-check, we also tried to fit the SEDs of all these dropouts assuming a low-$z$ solution, i.e. with a redshift space parameters ranging from 0.0 to 3.0 and allowing the same reddening interval. The best-fit is found, in each case, with a higher $\chi^2$ and a P(z) not well defined in the redshift interval explored (cf. Table \ref{physical_props}). 
We also used a prior in luminosity to check the consistency of our sample with previous studies at $z$$\sim$6. The prior was defined using the parametrization of the Schechter function published in \citet{Bouwens14}. The parameters space was defined as previously and the same templates library was used. Two candidates showed 
different best-fit photo-$z$'s, but the best-fit SED template was always found at $z>5.5$.

The UV luminosities were computed using the best-fit SEDs found by \textit{Hyperz} and the SFRs were computed from the UV luminosities using the relationship published in \citet{K98}. We corrected for the dust extinction of $L_{1500}$ following the \citet{Calzetti00} method. We noticed that the $L_{1500}$ is ranging from $\sim$2.5 to 29 $\times$10$^{41}$ergs s$^{-1}$ \AA $^{-1}$ and the SFRs from $\sim$26 to 300 M$_{\odot}$ yr$^{-1}$. These values are in good agreement with expectations for $z>$6 galaxies (e.g. \citealt{Schaerer10}) and with properties of previous samples at $z$$\sim$6-7 (\citealt{Zheng09}, \citealt{Curtis13}). 

\begin{deluxetable*}{ccccccccccc}
\tabletypesize{\scriptsize}
\tablecaption{Photometric properties of the $z$$\sim$6 galaxy candidates selected in this study}
\tablewidth{0pt}
\tablehead{
		& \multicolumn{4}{c|}{high-$z$}	& \multicolumn{3}{c|}{low-$z$}	& \multicolumn{2}{c}{Physical properties}	&  \\ \cline{2-10}
ID 		& $z_{phot}$	& $\chi^2$		&A$_v$	& 1$\sigma$ interval	& $z_{phot}$	& $\chi^2$		&A$_v$	& L$_{1500}$	& SFR  	 \\
 		& 			& 			&[mag]	& 				& 			& 			& [mag]	& [$\times$10$^{41}$erg/s/\AA]		& [M$_{\odot}$/yr]	
}
\startdata
z1\_GTC		&	6.5		& 0.6			& 0.8		&	6.3 - 6.5		& 1.1			& 2.2			& 0.0		& 28.8 		& 302.8					  \\
z2\_GTC		&	6.1		& 0.5			& 1.8		&	5.8 - 6.3		& 1.5			& 0.8			& 0.8		& 22.5 		& 236.0						  \\
z2\_HST		&	6.5		& 1.5			& 0.2		& 	6.0 - 7.2		& 1.4			& 4.6			& 0.0		& 4.5			& 47.0						  \\
z3\_HST		&	6.2		& 0.7			& 0.8		&	5.7 - 6.6		& 1.5			& 2.0			& 0.0		& 16.1		& 169.3						  \\
z4\_HST		&	6.4		& 0.3			& 0.8		&	5.7 - 7.8		& 1.5			& 1.3			& 0.0		& 11.6		& 121.8						  \\
z5\_HST		&	7.1		& 0.07		& 0.4		&	5.7 - 8.0		& 1.5			& 1.3			& 0.0		& 5.3			& 56.2						  \\
z6\_HST		&	7.1		& 0.02		& 0.2		& 	5.8 - 8.0		& 1.5			& 1.3			& 0.0		& 3.2			& 33.2						  \\
z7\_HST		&	5.6		& 0.2			& 0.8		&	4.1 - 6.2 		& 0.1			& 0.3			& 0.4		& 8.2			& 86.0						  \\
z8\_HST		&	6.2		& 0.3			& 0.8		&	5.3 - 7.4		& 1.5			& 0.6			& 0.0		& 8.5			& 89.5						  \\
z9\_HST		&	5.2		& 0.03		& 0.6		& 	0.0 - 6.1		& 0.9			& 0.1			& 2.2		& 2.5			& 26.2		  \\
\enddata
\tablecomments{\label{physical_props} Information given in this table : (1) ID, (2,3,4,5)  photo-$z$, $\chi_{red}$, $A_v$ and 1$\sigma$ confidence interval from the best SED-fit with a redshift parameter space ranging from $z\sim$0 to 8, (6,7,8)  photo-$z$, $\chi_{red}$ and $A_v$ for the best SED-fit assuming a low-$z$ solution (0$<z<$3), (9, 10) $L_{1500}$ deduced from best SED-fit and SFR computed using the \citet{K98} relation and corrected for dust extinction.
}
\end{deluxetable*}


The estimation of the photometric redshift generally implies the use of templates of nearby galaxies extrapolated to the very high-$z$. We made use of the large number of $z$$\sim$6 galaxies spectroscopically confirmed to check the reliability of our templates library and the SED-fitting method using data from  \citet{Toshikawa12}, \citet{Jiang13} and \citet{Willott13}. We matched photometric catalogs published in previous references to the quality of our dataset, more especially for the filters used here in the wavelength range covered by our data (from 0.4 to 1.6 $\mu$m). We re-computed photometric error bars of spectroscopically confirmed galaxies using the average noise measured in each of our images. In the case of a non-detection, we used the upper limits computed in our data.  The photometric redshift of the confirmed $z$$\sim$6 galaxies were computed with \textit{Hyperz} using the same templates as were used previously with the actual sample. Among the 50 galaxies used, 45 have a 1$\sigma$ confidence interval for the photometric redshift that includes the spectroscopic redshift. The mean absolute dilation to the spectroscopic redshift is $\sim$9\% with a standard deviation of $\sim$16\% showing that the method we used to estimate the photometric redshift of our candidates is efficient at $z$$\sim$6. We also used this spectroscopic sample to check the reliability of our color criteria, and demonstrated that all these spectroscopically confirmed galaxies would be selected using the criteria defined in Sec. \ref{LBG}.

\section{Environment of \textit{HFLS3}}
\label{environment}
In this Section, we describe the environment of the starburst galaxy \textit{HFLS3}, and discuss the possibility for an
overdensity associated with it.



\subsection{Faint $z$$\sim$6 candidates over the wide field}
\label{faint}

\begin{figure*}
\centering
\includegraphics[angle=0,scale=.65]{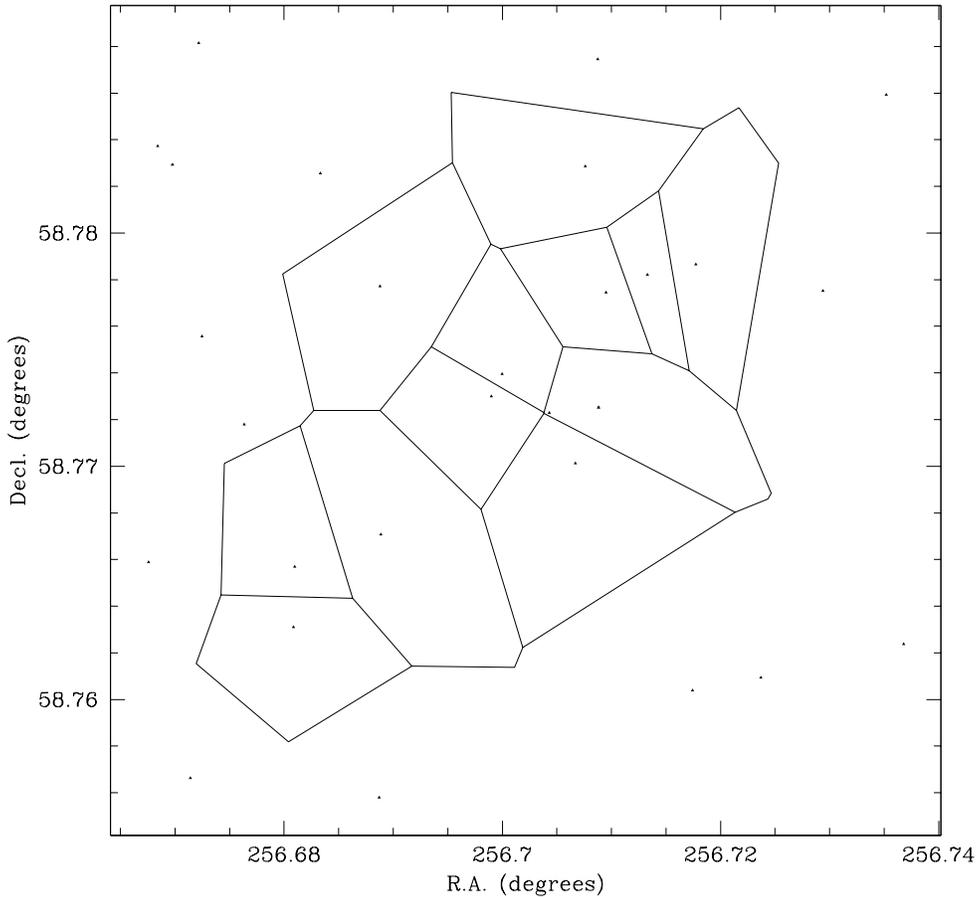}
\caption{\label{voronoi} Results of the Voronoi tessellation analysis following the method described in \citet{Ramella01} on the sample of faint sources described in \ref{faint}. The background threshold above which fluctuations are consistent with an over-dense field is estimated at 1.74 per Voronoi cell, which is not a significant
detection of an over-density in the imaging data. }
\end{figure*}

In order to select faintest objects at the redshift of \object{HFLS3},
we relaxed the detection constraints we imposed for our HST selection (eq. \ref{det_HST}) by requiring a detection $\ge$3$\sigma$ in F105W and F125W (instead of the 5$\sigma$ level required previously on F105W) and by using the F814W-F105W color criterion computed previously. Indeed by relaxing the detection level on the F105W image, we are not able to use the color criterion combining F625W and F814W, because of the shallower depth of the F625W. After visual inspection removing false detection as explained above, 25 objects satisfy this new selection function. In order to study if this sample of faint objects shows any evidence for an overdensity of galaxies associated with \object{HFLS3}, we applied the Voronoi tessellation method as described in \citet{Ramella01} and based on the "triangle" C code published by \citet{Shewchuk96}. The density threshold we used to distinguish background regions and fluctuations which are significant overdensities is estimated at 1.74 object per Voronoi cell. As seen in figure \ref{voronoi}, no over-dense region is highlighted by the new faint objects sample. 

\begin{figure*}
\centering
\includegraphics[angle=0,scale=.65]{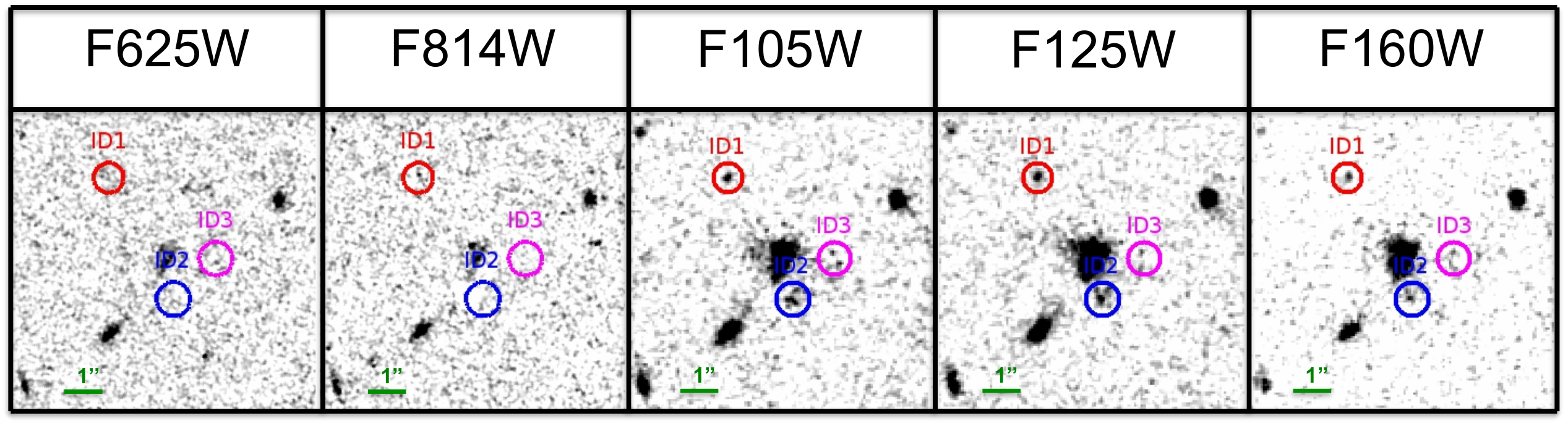}
\caption{\label{close_to_FLS3} Thumbnail images of the faint sources located around HFLS3. The size of each stamp is $\sim$8.3''$\times$8.3''. The lens system studied in \citet{Cooray14} is at the center of the field. The position of each faint candidate is display by a circle. }
\end{figure*}

However, we noticed that three sources close to \textit{HFLS3} (with 26.5$<$m$_{F105W}$$<$28.0) are not detected on ACS data (figure \ref{close_to_FLS3}). Their break between F814W and F105W is $<$1 mag and their NIR colors could be consistent with the Balmer break, therefore we cannot exclude a low-$z$ solution for these three objects. Assuming that these three sources are at the redshift of \textit{HFLS3}, the projected distance between them and the \textit{Herschel} starburst is less than 15 kpc. We also compared the overdensity of objects with brightness similar to these three objects over the entire field of view covered by HST. The mean density over the field is 0.03 object per arcsec$^2$ whereas it is 4$\times$ more around HFLS3, reinforcing the overdensity hypothesis close to the Herschel starburst. We applied the same SED-fitting method described above with the same library of templates. We first allowed a large range of redshift (0$<z<$8), but found that all these objects to be described  SEDs compatible with the SED of a $z$$\sim$6.3 galaxy. However, we have to keep in mind that the large error bars as well as the small break between F814W and F105W could not exclude a low-$z$ solution for all of them. Deep ACS and WFC3 data are needed to strongly increase the non-detection constraints in optical and reduce the error bars on the WFC3 photometry. 

\subsection{Discussion }

In order to constrain the size of a possible overdensity, we adopted the galaxy fluctuation parameter defined in \citet{Morselli14}:
\begin{equation}
\delta = \frac{\rho}{\dot{\rho}} - 1
\end{equation}
where $\rho$ is the number of objects selected in our survey and $\dot{\rho}$ the number of objects expected in a blank field covering the same area.

We then compared the number of objects selected in a luminosity range where the completeness is $\sim$100\% with the number of objects expected in the same interval in magnitude in a blank field. We computed the completeness level of the F105W image, where the $L_{1500}$ is estimated, by adding 10,000 sources per bin of 0.25 magnitude. We then applied the extraction parameters explained in section \ref{extraction} and compared the number of sources extracted with the number of objects added on the image. We showed that $\sim$100\% of our added objects are detected up to $m_{F105W}<$25.3. Three objects in our HST sample are brighter than this magnitude cut.
We used the UV luminosity function parameterization published in \citet{McLure09}, \citet{Su11}, \citet{Bouwens12} and \citet{Bouwens14} to estimate the number of $z$$\sim$6 galaxies expected in our HST survey with $m_{F105W}<$25.3 (cf. Table  \ref{parameterization}). The expected number of objects ranges from 0.5 to 1.0, therefore the over-density parameter ranges from 2.4 to 9.15, with 9.15 as a strict upper limit on the over-density given the cosmic variance and taking into account the contamination rate computed in section \ref{contaminants}. We emphasize that this result is
based on a small field-of-view and a small number of objects. \citet{Chapman09} found a similar value ($\delta\sim$2.5) for UV-selected galaxies over the GOODS-N field, and noticed that this value is lower than that found for SMGs in the same field ($\delta\sim$10). As shown by \citet{Robson14} there is no over-density of SMGs around \textit{HFLS3}. If an over-density occurs around \textit{HFLS3}, it is at a lower level than in the COSMOS AzTEC-3 field \citep{Capak11}  where they found an over-density factor of 11. 

We note that four galaxies among our two samples have high SFRs estimated from the SED models (with SFRs $>$ 100 M$_{\odot}$ yr$^{-1}$). 
These sources, however, are undetected in the SCUBA-2 850 $\mu$m image of this field \citep{Robson14}, with 
a noise level of $\sim$ 1.5 mJy/beam.  \citet{Riechers14} have studied with ALMA the center of the galaxy protocluster associated with 
the $z$=5.3 SMG AzTEC-3 \citep{Capak11}, but failed to detect a LBG at $z$=5.3 with a SFR $\sim$ 20 M$_{\odot}$ yr$^{-1}$ with ALMA
data reaching $\sim$ 0.15 mJy/beam (3$\sigma$) at 1 mm. Assuming typical T$_{dust}$ of 35K for the {\it HFLS3} LBGs, 
a colder dust temperature than \textit{HFLS3}, observed-frame 850$\mu$m lies on the Wien part of the SED and the $K$ correction from $z$=5.3 to $z$=6.3 remains positive. Scaling the SED and depths reached between ALMA at $z=5.3$ at 1 mm and  SCUBA-2 for galaxies at $z \sim 6.3$
at 850 $\mu$m, we find that the SCUBA-2 depth is only adequate to detect galaxies above 3 $\sigma$ 
with instantaneous SFRs $>$ 500  M$_{\odot}$ yr$^{-1}$. This ignores the cosmic microwave background (CMB) that
will absorb a few per cent of the sub-mm flux at $z$=6.3 compared to $z$=5.3. Therefore SCUBA-2 observations of the depth of the \citet{Robson14} data are unlikely to
detect galaxies with similar SFRs and SED properties as LBGs in the AzTEC-3 protocluster.

\subsection{Distribution in luminosity of the $z$$\sim$6 candidates }
\label{UV}

The redshift interval covered by our survey is sufficient to compute the UV luminosity function at $z$$\sim$6. This again allows a way to
study if there is an overdensity of objects at a given luminosity. 
Several methods have been developed to compute the UV LF at different redshifts (eg. \citealt{Bouwens12arxiv}, \citealt{Willott13}). 
We made use of the previous SED-fitting work to compute the number density of objects at $z$$\sim$6 as a function of the UV luminosity (see \citealt{Laporte15} for details). 

The effective surface explored by our HST survey is computed by masking the bright objects on the detection image. The comoving volume explored between $z\sim$5.5 and $z$$\sim$6.5 is $V\approx$10 980 Mpc$^3$. The number densities we found are shown on Table \ref{density} and the error bars take into account the cosmic variance. These numbers are consistent with previous findings in this redshift interval (e.g. \citealt{Willott13}, \citealt{Bouwens07}).

We adopted the Schechter parameterization \citep{Schechter} of the UV LF defined by:
\begin{equation}
\small
\Phi(M)=\Phi^{\star}\frac{\ln(10)}{2.5}\big(10^{-0.4(M-M^{\star})}\big)^{\alpha+1} \exp \big(-10^{-0.4(M-M^{\star})}\big)
\end{equation}
where $M^{\star}$, $\Phi^{\star}$ and $\alpha$ are the 3 Schechter parameters to be adjusted. 

\begin{figure*}
\centering
\includegraphics[angle=0,scale=.55]{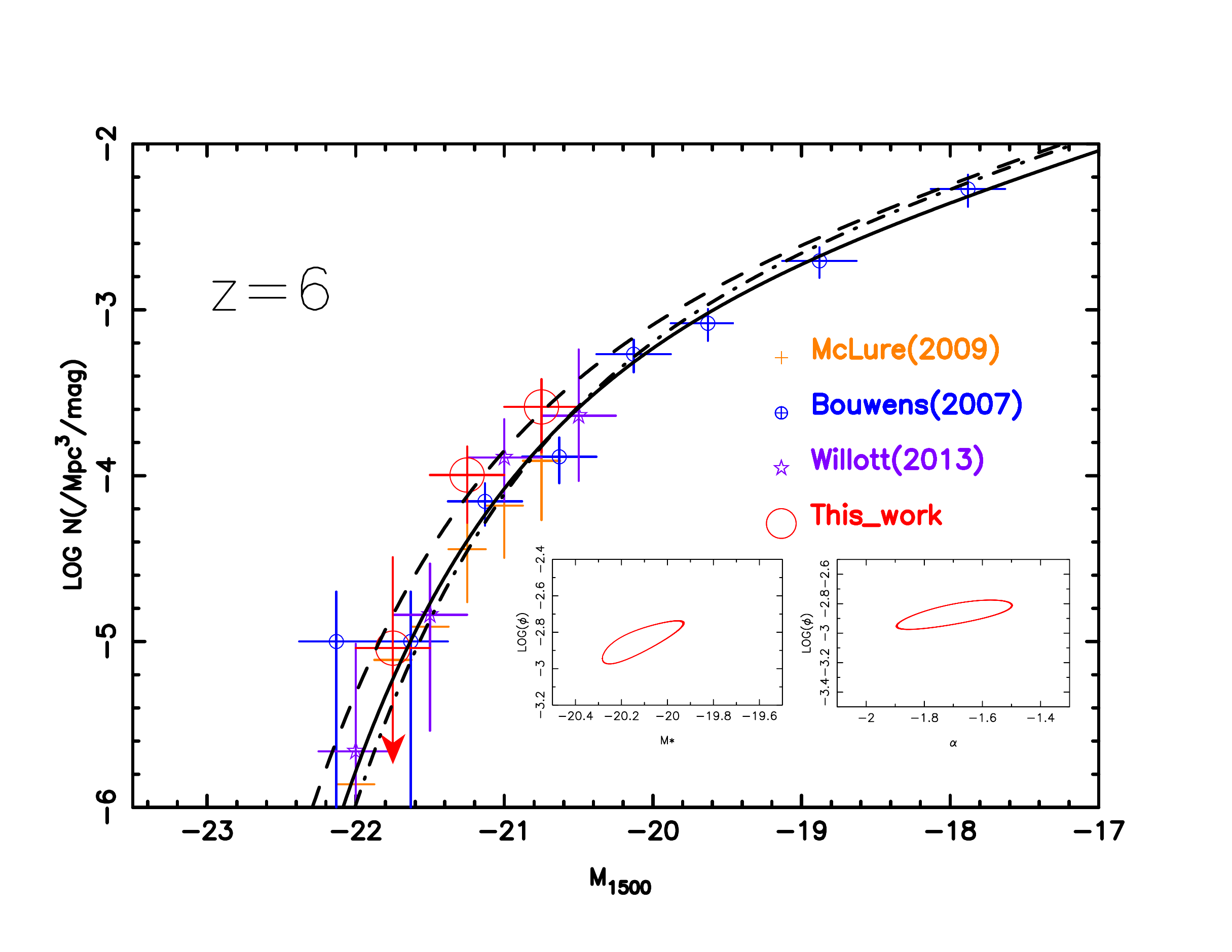}
\caption{\label{LF_plot} Shape of the UV luminosity function at $z$$\sim$6. Number densities from this study are plotted in red, previous finding are also plotted (\citealt{McLure09}, \citealt{Bouwens07} and \citealt{Willott13}). The parameterization computed from the study of HFLS3 environment is displayed by the black line, the other lines display the parameterization published in previous references (dashed line from \citealt{Bouwens12} and dotted-dashed line from \citealt{McLure09}). The 1$\sigma$ confidence intervals on each parameter are also shown on the lower panels.}
\end{figure*}

In order to show the influence of the densities computed in this study, we fitted the shape of the UV LF by combining them with the density from \citet{Bouwens14}  in the faint-end slope and in the bright-end with those published by \citet{Willott13}. We used a $\chi^2$ minimization with parameters ranging from 10$^{-5}$ to 10$^{-2}$ Mpc$^{-3}$.mag$^{-1}$,  -22.0 to -19.0 and  -2.2 to -1.1 respectively for $\Phi^{\star}$, $M^{\star}$ and $\alpha$. The following parameterization is found at $z$$\sim$6 : $M^{\star}$=-20.17$^{+ 0.35}_{-0.15}$, $\Phi^{\star}$=(1.21$^{+0.40}_{-0.19}$)$\times$10$^{-3}$Mpc$^{-3}$.mag$^{-1}$ and $\alpha$=-1.77$\pm$0.23. Error bars on each parameters are deduced from the 1$\sigma$ confidence interval. These parameters are in good agreement with previous results (e.g. \citealt{Bouwens12}, \citealt{Su11} and \citealt{McLure09} - Table \ref{parameterization}). The shape of the UV LF deduced from this study as well as the number densities we computed are plotted on Figure \ref{LF_plot}.
We find no evidence for an over-density of galaxies, above the field LF, at any of the luminosities probed by the data.

\begin{deluxetable}{ccc}
\tabletypesize{\scriptsize}
\tablecaption{Number density at $z$$\sim$6}
\tablewidth{0pt}
\tablehead{$M_{1500}$	&	$\Phi(M_{1500})$		&	d$\Phi$ \\
	& $\times$10$^{-4}$[Mpc$^{-3}$.mag$^{-1}$]	&	$\times$10$^{-4}$[Mpc$^{-3}$.mag$^{-1}$]	\\
}
\startdata
-21.75$\pm$0.250		&	0.09	&$^{+ 0.23}_{-0.09}$ \\
-21.25$\pm$0.250		&	1.01		& $\pm$0.49 \\
-20.75$\pm$0.500		&	2.60		& $\pm$1.22 \\
\enddata
\tablecomments{\label{density}
Number density computed following a method using the redshift probability distribution. Error bars included poisson uncertainties and cosmic variance computed from \citet{CV} }
\end{deluxetable}
\begin{deluxetable*}{cccc}

\tabletypesize{\scriptsize}
\tablecaption{UV LF parameterization at $z$$\sim$6}
\tablewidth{0pt}
\tablehead{Reference 	& $M^{\star}$	&	$\Phi^{\star}$		&	$\alpha$ \\
	& & $\times$10$^{-3}$[Mpc$^{-3}$.mag$^{-1}$]	&		\\
}
\startdata
This work			&	-20.16$^{+ 0.35}_{-0.15}$			&	1.36$^{+ 0.40}_{-0.19}$	& -1.70$\pm$0.23 \\
\citet{McLure09}	&	-20.04$\pm$0.12				& 1.80$\pm$0.50			& -1.71$\pm$0.11 \\
\citet{Su11}		& 	-20.25$\pm$0.23				& 1.77$^{+ 0.62}_{-0.49}$		& -1.87$\pm$0.14 \\
\citet{Bouwens12}	&	-20.37$\pm$0.30				& 1.4$^{+ 1.1}_{-0.6}$		& -1.73$\pm$0.22 \\
\citet{Bouwens14}	&	-20.93$\pm$0.25				& 0.49$^{+0.26}_{-0.17}$		& -1.85$\pm$0.10 \\
\enddata
\tablecomments{\label{parameterization} Parameterization of the Schechter function presented in this paper and published by other teams. }
\end{deluxetable*}

\section{Conclusions}
We have presented in this paper the results of the photometric analysis of the environment of a sub-millimeter starburst at $z$$\sim$6.34,
 combining both wide area ground-based data and high-level quality HST data
to explore a large range of luminosities. We applied the Lyman-break galaxy selection technique and found 10 galaxies that are at $z$$\sim$6. This sample
includes two from our ground-based search ($m_z$$\sim$25) and eight  more from our search with HST/WFC3 data (with $m_{F105W}$ in the range 24.5 to 26.5). The size of each sample is well-consistent with expectations from previous findings using blank field surveys and seems incompatible with an over-density of luminous ($m_{F105}$$<$25.9) galaxies. We used a standard method to estimate the photometric properties of each source and we used these SED-fitting results to compute the UV LF at $z$$\sim$6. The parameterization of the Schechter function we deduced ($M^{\star}$=-20.17$^{+ 0.35}_{-0.15}$, $\Phi^{\star}$=(1.21$^{+0.40}_{-0.19}$)$\times$10$^{-3}$Mpc$^{-3}$ and $\alpha$=-1.77$\pm$0.23) is in good agreement with previous finding in this redshift interval. 

We do not find any strong evidence for \textit{HFLS3} being a member of a proto-cluster of luminous galaxies as is the case of the well known SMG Aztec-3 \citep{Capak11}.
There the overdensity parameter was found to be about 11; in the case of  \textit{HFLS3}, we place an upper limit on the overdensity of $\approx$9 after taking into
account the cosmic variance of existing $z \sim 6$ LF measurements.
 The lack of a significat overdensity is also confirmed by a Voronoi tessellation analysis that included all the faint objects fulfilling the color-criteria defined for $z$$\sim$6 objects, but without a well-defined break between the optical ACS and near-IR WFC3 imaging data.

However, we noticed at least three faint objects within 3'' from \textit{HFLS3}. They are undetected in
ACS images and are detected on WFC3 data. 
If the redshift of these sources are confirmed at $z$$\sim$6, then the \textit{Herschel} starburst is located in an over-dense region composed by faint objects
but with an extent of 36 kpc. Deeper data combined to spectroscopic observations are needed to asses these conclusions. It is more likely that these
are associations with the merger system that might be triggering the starburst.

\begin{figure*}
\centering
   \begin{minipage}[b]{0.30\linewidth}
      \centering \includegraphics[scale=0.23]{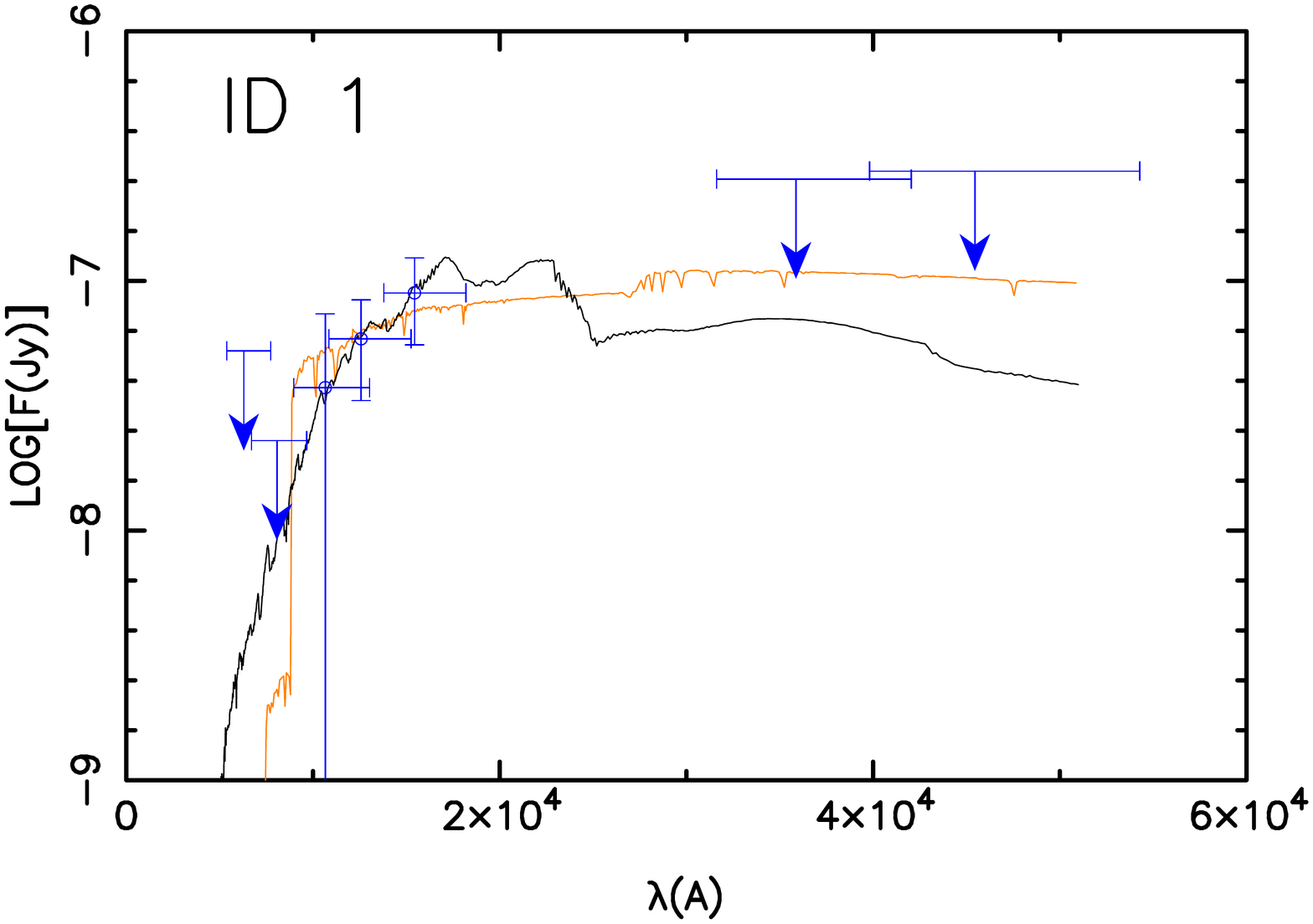}
   \end{minipage}
   \begin{minipage}[b]{0.30\linewidth}   
      \centering \includegraphics[scale=0.23]{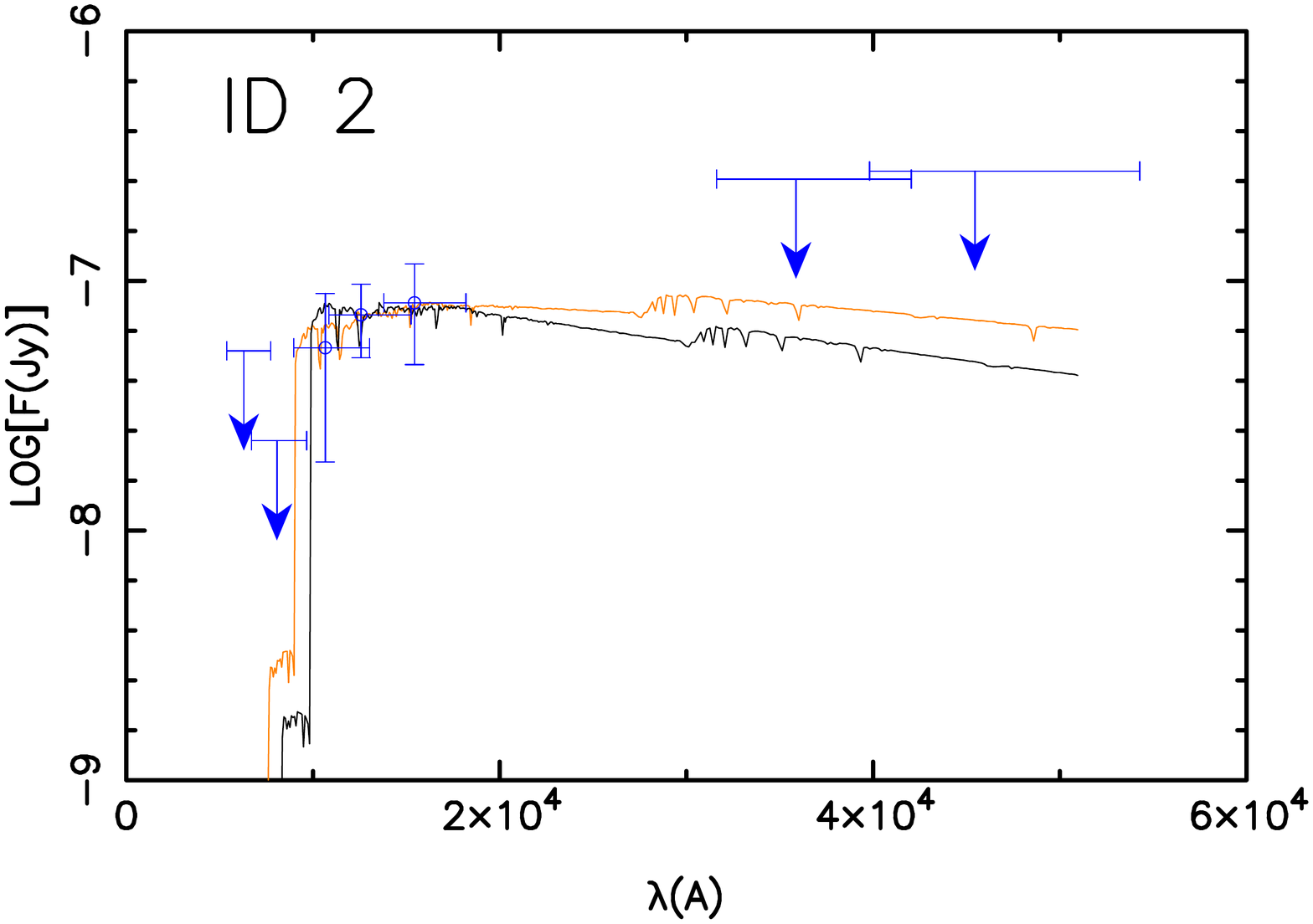}
   \end{minipage}
   \begin{minipage}[b]{0.30\linewidth}
      \centering \includegraphics[scale=0.23]{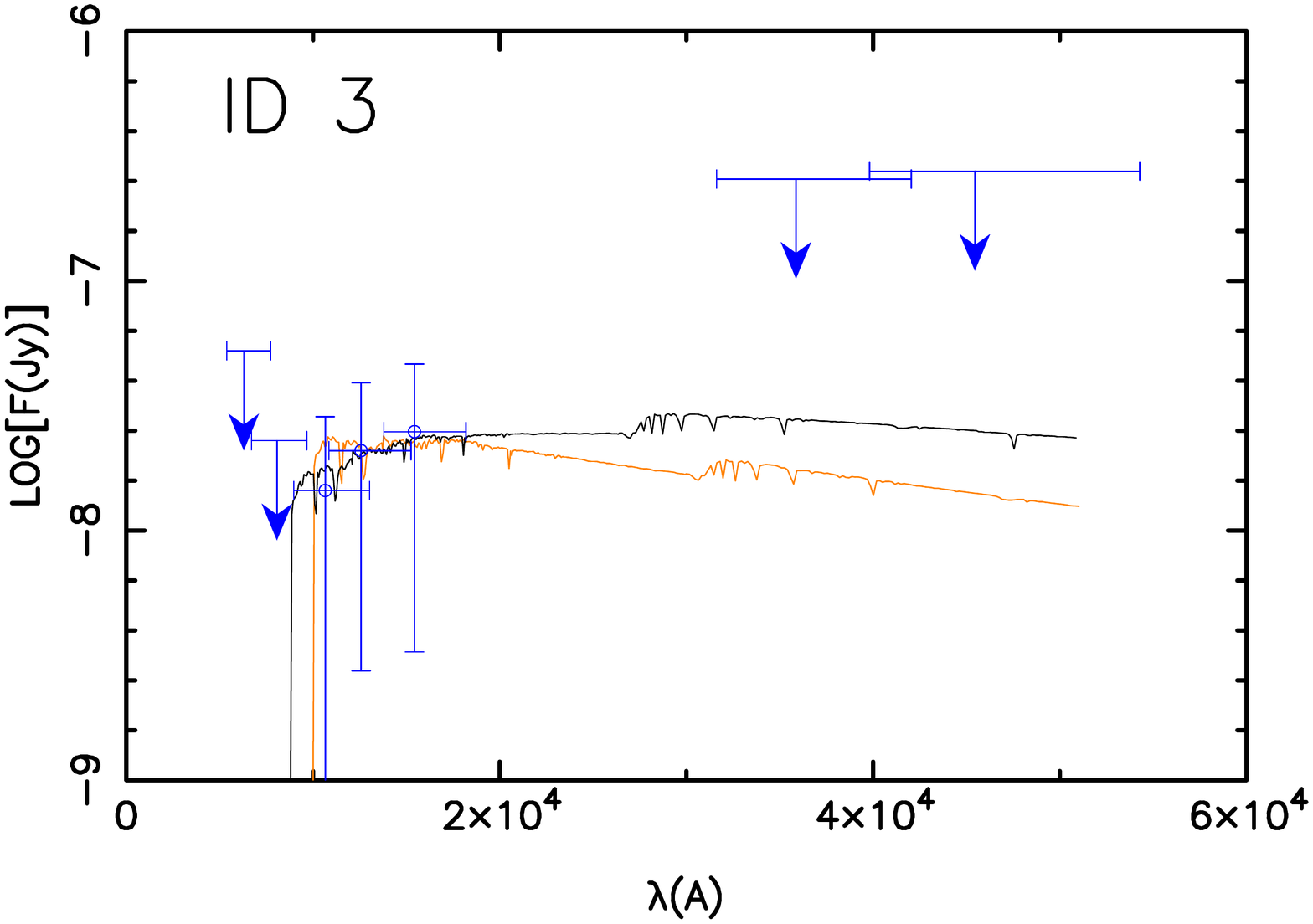}
   \end{minipage}\hfill
   \caption{SED-fitting of the 3 faint objects highlighted around \textit{HFLS3} using two assumptions on the redshift: (black line) allowing a large range of redshift 0$<z<$8 ($\chi^2_{red}\sim$0.05, 0.02, 0.01 respectively for ID1, ID2 and ID3) and (orange line) fixing the redshift at the redshift of \textit{HFLS3} ($\chi^2_{red}\sim$0.15,0.06, 0.01 respectively for ID1, ID2 and ID3). The blue dots show the photometry of these 3 faint objects (SExtractor MAG\_AUTO) and the 2$\sigma$ upper limits in case of non-detection. Error bars are computed using the noise measured in empty apertures around the objects.}
   \label{SED_overdensity}
\end{figure*}

\acknowledgments{
Financial support for this work was provided by the Spanish MINECO under projects AYA2010-21697-C05-04 and FIS2012-39162-C06-02, by the Chilean Basal-CATA PFB-06/2007 and CONICYT-Chile under the grant Gemini-CONICYT \#32120003,  by NASA through grant HST-GO-13045 from the Space Telescope Science Institute, which is operated by Associated Universities for Research in Astronomy, Inc., under NASA contract NAS 5-26555. Additional support for AC, WO, JC, JLW, and CMC was from NSF with AST-1313319. Dark Cosmology Centre is funded by the Danish National Research Foundation (JLW). SO acknowledges support from the Science and Technology Facilities Council [grant number ST/I000976/1]. DR acknowledges support from the Science and Technology Facilities Council (ST$/$K00106X$/$1). RJI acknowledges support from the European Research Council in the form of Advanced Grant, COSMICISM, 321302.SPIRE has been developed by a consortium of institutes led by Cardiff Univ. (UK) and including: Univ. Lethbridge (Canada); NAOC (China); CEA, LAM (France); IFSI, Univ. Padua (Italy); Stockholm Observatory (Sweden); Imperial College London, RAL, UCL-MSSL, UKATC, Univ. Sussex (UK); and Caltech, JPL, NHSC, Univ. Colorado (USA). This development has been supported by national funding agencies: CSA (Canada); NAOC (China); CEA, CNES, CNRS (France); ASI (Italy); MCINN (Spain); SNSB (Sweden); STFC, UKSA (UK); and NASA (USA). We thank Jos\'e Acosta-Pulido for advice on the LIRIS observations and help with the LIRIS data reduction package.This work is based on observations made with the Gran Telescopio Canarias (GTC), instaled in the Spanish Observatorio del Roque de los Muchachos of the Instituto de Astrof\'isica de Canarias, in the island of La Palma. The William Herschel Telescope is operated on the island of La Palma by the Isaac Newton Group in the Spanish Observatorio del Roque de los Muchachos of the Instituto de Astrofísica de Canarias.The data presented in this paper will be released through the Herschel Database in Marseille HeDaM (hedam.oamp.fr/HerMES)}


{\it Facilities:} \facility{GTC (OSIRIS)}, \facility{HST (ACS)}, \facility{HST (WFC3)}, \facility{Spitzer (IRAC)}, \facility{WHT (LIRIS)}.

\clearpage





\end{document}